\documentclass[conference]{IEEEtran}
\IEEEoverridecommandlockouts

\usepackage{cite}
\usepackage{amsmath,amssymb,amsfonts}
\usepackage{algorithm}
\usepackage{algpseudocode}
\usepackage{graphicx}
\usepackage{textcomp}
\usepackage{xcolor}

\usepackage{tabularx}
\usepackage{booktabs}
\usepackage{multirow}

\usepackage[utf8]{inputenc}    
\usepackage[T1]{fontenc}       

\usepackage{ragged2e}   
\usepackage{caption}    
\usepackage{array}      

\newcolumntype{Y}{>{\RaggedRight\arraybackslash}X}

\def\BibTeX{{\rm B\kern-.05em{\sc i\kern-.025em b}\kern-.08em
    T\kern-.1667em\lower.7ex\hbox{E}\kern-.125emX}}
\begin{document}

\title{Hybrid Decoding: Rapid Pass and Selective Detailed Correction for Sequence Models}


\author{
  \IEEEauthorblockN{Yunkyu Lim, Jihwan Park, Hyung Yong Kim, Hanbin Lee, Byeong-Yeol Kim
  }
  \IEEEauthorblockA{42dot Inc., Republic of Korea\\
  yunkyu.lim@42dot.ai}
}

\maketitle

\begin{abstract}
Recently, Transformer-based encoder-decoder models have demonstrated strong performance in multilingual speech recognition. However, the decoder's autoregressive nature and large size introduce significant bottlenecks during inference. Additionally, although rare, repetition can occur and negatively affect recognition accuracy. To tackle these challenges, we propose a novel Hybrid Decoding approach that both accelerates inference and alleviates the issue of repetition. Our method extends the transformer encoder-decoder architecture by attaching a lightweight, fast decoder to the pretrained encoder. During inference, the fast decoder rapidly generates an output, which is then verified and, if necessary, selectively corrected by the Transformer decoder. This results in faster decoding and improved robustness against repetitive errors. Experiments on the LibriSpeech and GigaSpeech test sets indicate that, with fine-tuning limited to the added decoder, our method achieves word error rates comparable to or better than the baseline, while more than doubling the inference speed.







\end{abstract}


\section{Introduction}
Recent advances in automatic speech recognition (ASR) have been driven by two complementary trends: improvements in transcription accuracy and increases in inference speed \cite{AED, LAS, two-pass, ctc, rnnt}. Architecturally, transformer encoder-based models have become dominant in the field~\cite{transformer,fastconformer}. Innovations in loss functions-including Connectionist Temporal Classification (CTC)\cite{ctc}, RNN-Transducer (RNN-T)\cite{rnnt}, CR-CTC~\cite{cr-ctc}, and Token-and-Duration-Transducer (TDT)~\cite{tdt}—have provided alternative frameworks that improve both learning efficiency and recognition performance.

In parallel, data-centric techniques such as self-supervised learning (SSL)\cite{ASBERT, SelftrainingASR, wav2vec2} have delivered substantial gains by enabling pretraining on large volumes of unlabeled speech data. The combined effect of architectural and data-driven innovations has led to ASR systems capable of remarkable performance in multilingual settings \cite{xlsr}.

With the emergence of increasingly agentic artificial intelligence, the need for robust and efficient speech recognition as a foundation for natural human–agent interaction has grown even more important. In these interactive settings, models are often expected to support multiple speech-related tasks, such as speaker diarization and emotion recognition, alongside standard transcription. Supporting these diverse objectives within a unified model architecture comes with significant computational and architectural challenges.

Transformer-based attention encoder-decoder (AED) architectures, exemplified by Whisper~\cite{Whisper} and Canary~\cite{canaryflash}, have demonstrated strong multitask and multilingual abilities using prompt-conditioned decoding. However, this flexibility has been accompanied by a dramatic increase in decoder parameter counts and inference costs. The fully autoregressive nature and substantial size of transformer decoders lead to notable latency, especially compared to more efficient models based on CTC or RNN-T~\cite{canaryflash,distill-whisper}.

To mitigate these challenges, recent research has explored various model optimizations. Canary-1B-Flash~\cite{canaryflash}, for example, achieves a significant speedup by reducing decoder layers, redistributing computational effort to the encoder. Similarly, knowledge distillation and speculative decoding~\cite{distill-whisper,speculative} have yielded additional gains. Nonetheless, a considerable efficiency gap remains between AED and CTC/RNN-T architectures, particularly as output sequence lengths increase~\cite{canaryflash}.

In this work, we propose a novel hybrid decoding approach designed to address the fundamental limitations of inference speed and robustness against repetitive errors in AED-based ASR models. The key insight is that the primary bottleneck lies in the transformer decoder's requirement to generate each token sequentially, precluding parallelism and compounding latency for long utterances. Our method extends the standard AED framework by attaching a lightweight fast decoder to the pretrained encoder. During inference, this fast decoder quickly generates a reference sequence, which is then selectively reviewed and corrected by the transformer decoder using our hybrid decoding algorithm. Table~\ref{tab:hybrid-decoding-examples} presents an example illustrating the operation of the proposed algorithm.

This two-pass hybrid decoding strategy substantially reduces the number of required sequential forward steps, resulting in significant improvements in inference speed without compromising recognition accuracy. We evaluate our approach on both the in-domain LibriSpeech~\cite{librispeech} and out-of-domain GigaSpeech~\cite{gigaspeech} test sets. The results demonstrate that, with finetuning limited to the fast decoder, our method achieves word error rates (WERs) comparable to or better than the baseline, while more than doubling the inference speed.

\begin{table*}[t]
\centering
\small
\setlength{\tabcolsep}{4pt}
\renewcommand{\arraystretch}{1.0} 

\caption{Examples of error correction by hybrid decoding. Each row illustrates a specific error type (insertion, deletion, substitution) in the fast decoder output and demonstrates how the proposed two-pass algorithm identifies and corrects it using the transformer decoder. \texttt{Ref} denotes the initial hypothesis from the fast decoder, \texttt{Verif} is the transformer decoder's teacher-forced prediction, \texttt{Diff} shows the first divergence point, \texttt{Patch} is the generated correction, \texttt{Range} specifies the replaced segment, \texttt{Result} is the final output, and \texttt{Steps} indicates the reduction in transformer decoder forward steps from baseline to the proposed method. This table highlights the localized and efficient correction capability of the hybrid decoding strategy.}
\label{tab:hybrid-decoding-examples}
\begin{tabularx}{\textwidth}{@{}lY@{}}
\toprule
\textbf{Type} & \textbf{Correction Pipeline} \\
\midrule
\textbf{Insertion} &
\begin{tabular}[t]{@{}l@{}}
\textbf{\ttfamily Ref\ \ \ \ }: \textbf{we} token by which i shall discover it\\
\textbf{\ttfamily Verif\ \ }: \textbf{wh'}oken by which i shall discover it\\
\textbf{\ttfamily Diff\ \ \ }: idx 0 (\texttt{we} $\rightarrow$ \texttt{wh})\\ 
\textbf{\ttfamily Patch\ \ }: where the token\\
\textbf{\ttfamily Range\ \ }: we token\\
\textbf{\ttfamily Result\ }: where the token by which i shall discover it\\
\textbf{\ttfamily Steps\ \ }: 17 $\rightarrow$ 8 
\end{tabular} \\[0.4em] 
\midrule
\textbf{Deletion} &
\begin{tabular}[t]{@{}l@{}}
\textbf{\ttfamily Ref\ \ \ \ }: the girl who breaks the rules \textbf{haves} to be punished\\
\textbf{\ttfamily Verif\ \ }: the girl who breaks the rules \textbf{hasvees} to be punished\\
\textbf{\ttfamily Diff\ \ \ }: idx 13 (\texttt{ha} $\rightarrow$ \texttt{has})\\ 
\textbf{\ttfamily Patch\ \ }: has to be punis\\
\textbf{\ttfamily Range\ \ }: haves to be punis\\
\textbf{\ttfamily Result\ }: the girl who breaks the rules has to be punished\\
\textbf{\ttfamily Steps\ \ }: 22 $\rightarrow$ 8 
\end{tabular} \\[0.4em]
\midrule
\textbf{Substitution} &
\begin{tabular}[t]{@{}l@{}}
\textbf{\ttfamily Ref\ \ \ \ }: i dunno muttered dick and \textbf{our} men can't be sure\\
\textbf{\ttfamily Verif\ \ }: i dunno muttered dick and \textbf{the} men can't be sure\\
\textbf{\ttfamily Diff\ \ \ }: idx 11 (\texttt{our} $\rightarrow$ \texttt{the})\\ 
\textbf{\ttfamily Patch\ \ }: the men can't be\\
\textbf{\ttfamily Range\ \ }: our men can't be\\
\textbf{\ttfamily Result\ }: i dunno muttered dick and the men can't be sure\\
\textbf{\ttfamily Steps\ \ }: 20 $\rightarrow$ 8 
\end{tabular} \\
\bottomrule
\end{tabularx}
\end{table*}

\section{Related Work}
Canary-Flash~\cite{canaryflash} has demonstrated that the bulk of inference overhead in attention-based encoder-decoder speech models stems from the autoregressive cross-attention decoder. To mitigate this, the model architecture can be rebalanced by expanding the encoder while reducing the number of decoder layers, thus shifting computational demand toward the encoder. This adjustment enables significant inference speedups—up to 3$\times$—without sacrificing accuracy or requiring changes to the dataset or convergence conditions. However, when compared to CTC~\cite{ctc} or transducer-based architectures~\cite{tdt}, a considerable gap in inference efficiency remains, and this disparity grows as the number of output tokens increases. Our proposed two-pass hybrid decoding method reduces the forward steps of the transformer decoder, delivering more than a twofold speedup over even optimized encoder-intensive models like Canary-Flash, while ensuring equivalent recognition performance.

Distil-Whisper~\cite{distill-whisper} approach enhances inference latency by distilling large Whisper~\cite{Whisper} models using extensive, pseudo-labeled, open-source datasets. In this framework, the student model is created by reducing the number of decoder layers from 32 to 2 and training on high-quality pseudo-labels selected with an optimized WER threshold. This process yields up to a 5.8$\times$ improvement in inference speed with only a slight reduction in performance, and achieves approximately $2\times$ faster inference at equivalent WER when paired with the teacher, original whisper, model for speculative decoding~\cite{speculative}. However, when the number of decoder layers is already small (e.g., four layers in Whisper-Turbo), the speedup from distillation becomes negligible. In contrast, our proposed method introduces only a lightweight module to a pretrained model, yet consistently achieves a $2\times$ improvement in inference efficiency without compromising recognition accuracy, even for architectures with four decoder layers.

Two-Pass End-to-End Speech Recognition~\cite{two-pass} introduces an architecture in which an RNN-T decoder~\cite{rnnt} and a LAS~\cite{LAS} decoder share a common encoder. While both this previous work and our method employ two sequential decoding stages, there are notable differences. First, whereas the earlier framework utilizes an LSTM layer in the second pass, our approach incorporates a transformer layer, which entails significantly higher computational requirements and thus necessitates greater emphasis on optimizing inference speed. Second, whereas the prior work prioritizes improving the rescoring capability of the second-pass model—employing techniques such as minimum word error rate training—our method adopts an explicit algorithmic process that iteratively verifies and refines the first-pass outputs to ensure alignment with the second-pass predictions.

\section{Methods}
We propose two key components to improve the inference efficiency of AED-based ASR:

1. The introduction of a fast decoder to reduce the number of autoregressive forward steps required by the transformer decoder

2. A novel, simple yet highly effective two-pass hybrid decoding algorithm.

\subsection{Fast Decoder}
To alleviate the computational burden of the transformer decoder, we introduce a lightweight yet high-performing fast decoder. It rapidly generates a hypothesis for the entire utterance, which serves as a reference sequence. The transformer decoder then selectively performs detailed corrections only on the erroneous segments of this reference, thereby significantly reducing the number of required forward passes. Achieving optimal performance requires a fast decoder that is both computationally efficient and highly accurate.

Previous research~\cite{fastconformer} on CTC- and transducer-based speech recognition has demonstrated that decoders constructed with convolutional or recurrent networks can deliver strong performance with relatively few parameters. Among these, the token-and-duration transducer (TDT)~\cite{tdt} achieves a good balance between recognition accuracy and inference speed by jointly predicting both the token and its duration. Motivated by these results, we adopt the architecture proposed in TDT as our fast decoder.

In addition, ASBERT~\cite{ASBERT} has shown that the final embeddings from an encoder trained on ASR tasks encode linguistic features that are suitable even as pseudo labels. Building on this insight, we perform fine-tuning exclusively on the fast decoder, keeping the backbone AED model frozen. This strategy dramatically reduces the resources required for obtaining an effective first-pass model, since only the fast decoder needs to be finetuned.

\subsection{Hybrid Decoding Algorithm}
We propose a hybrid decoding strategy that leverages a fast decoder to generate an initial reference sequence, followed by a selective second-pass verification and correction step using a transformer decoder. The algorithm iteratively verifies the reference and corrects only erroneous segments, significantly reducing redundant computation.

Let $\mathbf{y}^{\mathrm{ref}} = [y_0^{\mathrm{ref}}, \ldots, y_{N-1}^{\mathrm{ref}}]$ denote the reference sequence generated by the fast decoder, and let $\mathcal{T}(\cdot)$ denote the transformer decoder. In this context, the decoder emits a special token denoted as \texttt{eos} (end-of-sentence) to indicate the termination of the sequence. The model is trained such that the eos token only appears at the final position of an output sequence and never elsewhere. Therefore, a decoding process is considered complete if and only if the eos token has been generated.

\begin{algorithm}[htbp]
\caption{Hybrid Decoding}
\begin{algorithmic}[1]
\Require Sequence $\mathbf{x}$, reference sequence $\mathbf{y}^{\mathrm{ref}}$ (from fast decoder), transformer decoder $\mathcal{T}$, minimum patch verification length $K$
\Ensure Corrected output sequence
\State Initialize prompt state for transformer decoder using $\mathbf{x}$
\While{True}
    \State Run $\mathcal{T}$ in teacher-forcing mode on $\mathbf{y}^{\mathrm{ref}}$ to obtain predicted sequence $\mathbf{y}^{\mathrm{tf}}$ and an indicator \texttt{eos} that is true if and only if the eos token is generated
    \State Compute $i^* = \min \{\, i \mid i \geq |\mathbf{y}^{\mathrm{ref}}|$ or $i \geq |\mathbf{y}^{\mathrm{tf}}|$ or $y^{\mathrm{ref}}_i \neq y^{\mathrm{tf}}_i \,\}$
    \If{$i^* \geq |\mathbf{y}^{\mathrm{ref}}|$ and \texttt{eos}}
        \State \Return $\mathbf{y}^{\mathrm{ref}}$
    \ElsIf{$i^* \geq |\mathbf{y}^{\mathrm{ref}}|$ and not \texttt{eos}}
        \State Let $\mathbf{c} \gets \mathbf{y}^{\mathrm{ref}}$ \Comment{Current verified reference as context}
        \State Initialize an empty list $\mathbf{s}_{\mathrm{new}}$
        \For{$k = 1$ \textbf{to} $K$}
            \State Autoregressively decode the next token $\hat{y}$ with $\mathcal{T}$ using context $\mathbf{c}$
            \State Append $\hat{y}$ to both $\mathbf{c}$ and $\mathbf{s}_{\mathrm{new}}$
            \If{$\hat{y} = \texttt{eos}$}
                \State \textbf{break}
            \EndIf
        \EndFor
        \State Update $\mathbf{y}^{\mathrm{ref}} \leftarrow \mathbf{y}^{\mathrm{ref}} \mathbin\Vert \mathbf{s}_{\mathrm{new}}$
        \State \Return $\mathbf{y}^{\mathrm{ref}}$
    \Else
        \State Confirmed prefix: $\mathbf{y}^{\mathrm{ref}}_{0:i^*}$, set as initial context
        \State Generate new patch $\mathbf{p}$ by running $\mathcal{T}$ \textbf{autoregressively} from position $i^*$ for at least $K$ steps or until \texttt{eos} (eos is generated)
        \State Locate the first occurrence of $\mathbf{p}$'s last token within the next $2|\mathbf{p}|$ tokens in $\mathbf{y}^{\mathrm{ref}}$ starting at $i^*$
        \State Replace the corresponding segment of $\mathbf{y}^{\mathrm{ref}}$ with $\mathbf{p}$
    \EndIf
\EndWhile
\end{algorithmic}
\end{algorithm}

The hybrid decoding algorithm begins by using the fast decoder to generate an initial candidate (reference) sequence. This candidate is then passed to the transformer decoder, which runs in teacher-forcing mode to verify the sequence against the model's internal predictions. The index of the first difference $i^*$ between the reference sequence and the transformer decoder's output is identified as
\begin{equation}
i^* = \min \left\{\, i \in \mathbb{N} \;\middle|\; i \ge |\mathbf{y}^{\mathrm{ref}}| \;\text{or}\; i \ge |\mathbf{y}^{\mathrm{tf}}| \;\text{or}\; y_i^{\mathrm{ref}} \neq y_i^{\mathrm{tf}} \,\right\}
\label{eq:first_diff}
\end{equation}
If no discrepancies are found and the output sequence from the transformer decoder is completed (that is, the eos token is produced), decoding terminates successfully. Otherwise, if the sequence is not yet completed, the decoder continues autoregressively for at most $K$ steps to complete the sequence. This step serves as a corrective mechanism for possible boundary mismatches: the fast decoder may have produced a reference sequence that is too short due to deletion errors, or the transformer decoder may generate repetitive tokens that prevent proper sequence termination. By allowing up to $K$ extra steps, the hybrid decoding algorithm can resolve these discrepancies by either appending missing tokens or limiting repetitive token generation by the decoder.

If a mismatch is detected at position $i^*$, a correction patch $\mathbf{p} = [p_0, \dots, p_{l-1}]$ is generated by the transformer decoder, which resamples a contiguous segment beginning at $i^*$. The patch length $l$ is at most $K$, or until the decoder produces an \texttt{eos} token. The range to be replaced in the reference sequence is determined by greedily searching for the first occurrence of the patch's last token within the next $2|p|$ tokens in $\mathbf{y}^{\mathrm{ref}}$ after position $i^*$. Specifically, the end index $j^*$ of the replaced segment is computed as follows:
\begin{equation}
j^* =
\begin{cases}
\min\{\, j \mid j \geq i^*,\, y_j^{\mathrm{ref}} = p_{l-1} \}, & \text{if such $j$ exists} \\
0, & \text{otherwise}
\end{cases}
\end{equation}
Here, $p_{l-1}$ denotes the last token of the patch, and $|p|$ is the length of the patch. The search is restricted to the next $2|p|$ tokens in $\mathbf{y}^{\mathrm{ref}}$ starting from $i^*$, in order to avoid errors that may arise when the last token appears multiple times in the reference sequence. This step accommodates insertion, deletion, and substitution errors in the draft. While this method uses a greedy search and does not guarantee the globally optimal solution for reference update, we empirically found that this simple heuristic is highly effective in practice.

This selective correction mechanism balances correctness and computational efficiency, as it avoids unnecessary recomputation of already-correct subsequences when the initial reference closely matches the target output.

The choice of $K$ presents a trade-off. If $K$ is too small, consecutive errors made by the fast decoder may require repeated verifications, and deletion errors that exceed $K$ near the end of decoding may not be correctable. Conversely, if $K$ is too large, the decoder may include superfluous tokens that do not require correction, leading to unnecessary latency. We empirically determined the optimal value of $K$ through ablation studies.

\begin{table}[tbp]
\centering
\begin{tabular}{lcc}
\toprule
Component           & Architecture   & \# Parameters (M) \\
\midrule
Encoder             & Transformer    & 810.0 \\
Main decoder        & Transformer    & 72.8  \\
Fast decoder        & TDT            & 13.3  \\
\bottomrule
\end{tabular}
\caption{Breakdown of model parameter counts (in millions) for each component of the hybrid decoding system: the transformer encoder, the main (transformer) decoder, and the fast decoder (TDT).}
\label{tab:param_detail}
\end{table}

\section{Experiments}
To evaluate the effectiveness and robustness of our proposed method, we used LibriSpeech~\cite{librispeech} as the in-domain dataset and GigaSpeech~\cite{gigaspeech} as the out-of-domain dataset. LibriSpeech consists of 960 hours of read audiobooks, whereas GigaSpeech is compiled from a variety of domains, including audiobooks, YouTube, and podcasts. For the AED backbone model, we utilized the Canary-1b-Flash~\cite{canaryflash} model publicly released by NVIDIA NeMo framework~\cite{NeMo}. Finetuning of the fast decoder was also conducted using the same framework. Word error rate (WER) was measured without consideration of capitalization and punctuation, and latency was reported as the average over five runs on an NVIDIA V100 GPU.

\begin{table*}[tbp]
\centering
\begin{tabular}{lccccccccc}
\toprule
\multirow{2}{*}{Decoding method} & \multicolumn{3}{c}{WER (\%, lower is better)} &  & \multicolumn{3}{c}{Latency (ms, lower is better)} \\
\cmidrule{2-4} \cmidrule{6-8}
& test\_clean & test\_other & gigaspeech & & test\_clean & test\_other & gigaspeech \\
\midrule
Transformer          & 1.63 & 3.08 & 10.13 && 196 & 177 & 194 \\
TDT          & 2.11 & 3.76 & 14.55 && 45  & 41  & 43 \\
Hybrid ($K$=1)  & 1.63 & 3.08 & 10.09 && 58  & 56  & 78 \\
Hybrid ($K$=3)  & 1.63 & 3.08 & 10.07 && 59  & 56  & 79 \\
Hybrid ($K$=5)  & 1.63 & 3.08 & 10.08 && 60  & 58  & 85 \\
Hybrid ($K$=7)  & 1.63 & 3.08 & 10.09 && 62  & 61  & 91 \\
Hybrid ($K$=9)  & 1.63 & 3.08 & 10.10 && 62  & 62  & 96 \\
\bottomrule
\end{tabular}
\caption{Comparison of word error rate (WER) and decoding latency in milliseconds (ms) across LibriSpeech test clean, test other, and GigaSpeech test sets for the baseline transformer decoder, the fast decoder (TDT), and various hybrid decoding configurations with different patch sizes \(K\). Results demonstrate that hybrid decoding achieves equivalent or better WER while substantially reducing latency compared to the fully autoregressive baseline, validating its effectiveness on both in-domain and out-of-domain evaluation data.}
\label{tab:wer_latency_summary}
\end{table*}

\subsection{Results on fast decoder}
For the fast decoder used in hybrid decoding, we adopted the TDT architecture from ~\cite{tdt}. The duration range of TDT was set to 0–4, with $\sigma = 0.05$ and $\omega = 0.1$. The prediction network consists of two RNN layers with a dimensionality of 640, and the joint network is a linear layer with a dimension of 640. The total number of parameters after adding the fast decoder to the backbone model is summarized in Table~2.

\begin{figure*}[tbp]
\centering
\includegraphics[width=0.95\linewidth]{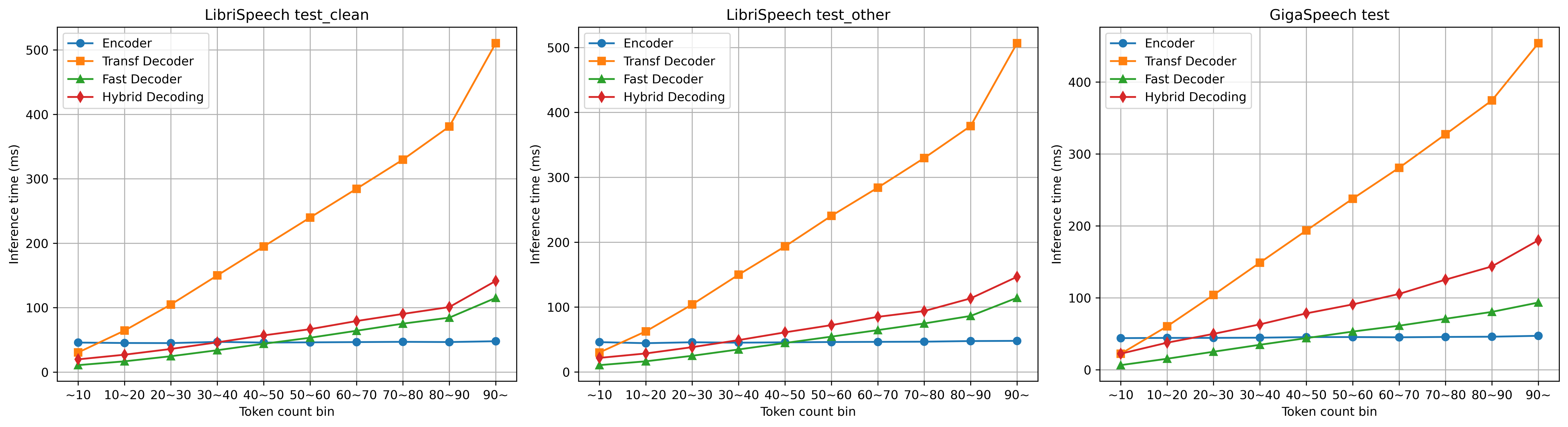}
\caption{Decoder inference time (ms) as a function of output token length quantiles (grouped in bins of 10 tokens) for (a) LibriSpeech test clean, (b) LibriSpeech test other, and (c) GigaSpeech test. The plots compare the latency of the transformer decoder, fast decoder, encoder module, and the proposed hybrid decoding \(K=3\). Results indicate that the latency of the baseline transformer decoder increases sharply with output length, while the hybrid approach maintains much lower and more stable inference times, especially for longer utterances.}
\label{fig:token_bin_inference_time}
\end{figure*}

We trained only the fast decoder on LibriSpeech data, keeping the AED modules of the backbone model frozen, and evaluated performance on both in-domain and out-of-domain data. Finetuning was performed on an A100 GPU using the AdamW optimizer with a global batch size of 2048, a learning rate of $1 \times 10^{-4}$, and for 7,000 steps. The results are shown in Table~3. With a total training time of only 64 GPU-hours, we obtained a fast decoder that delivers comparable word error rates on both LibriSpeech test clean and test other. On the out-of-domain GigaSpeech set, it achieved a meaningful word error rate of 14.55\%. Due to its lightweight architecture compared to the transformer-based decoder, the average decoder latency was greatly reduced—this corresponds to a minimum speedup of approximately 3.9 times.

\subsection{Results on hybrid decoding}
Hybrid decoding operates by treating the fast decoder output as a reference sequence, and repeatedly applying verification and correction using the transformer decoder. The parameter $K$ determines the maximum patch size, i.e., the number of tokens to be corrected in each correction step; experiments were performed with $K=1,3,5,7,9$. Table~\ref{tab:wer_latency_summary} summarizes WER and decoding latency for the transformer-based decoder, TDT, and hybrid decoding methods on LibriSpeech test clean, test other, and GigaSpeech.

On the in-domain LibriSpeech task, both the transformer decoder and all hybrid decoding variants achieved the lowest WER (1.63\% on test clean and 3.08\% on test other), indicating that the hybrid method successfully corrects first-pass errors. By comparison, TDT resulted in higher error rates (2.11\% and 3.76\%). The fast decoder was fine-tuned only on LibriSpeech, while the backbone model was trained on more diverse data; thus, the fast decoder exhibited larger errors on the out-of-domain GigaSpeech set. Nevertheless, hybrid decoding still delivered lower WER than the baseline on GigaSpeech, showing strong robustness.

When $K = 1$, no redundant correction is introduced, resulting in the lowest latency. However, in cases where consecutive token recognition fails, repeated verifications are triggered, making the latency comparable to the case of $K = 3$. As $K$ increases, the number of tokens that do not require correction in patch generation also increases, leading to a gradual rise in latency.

Further analysis on the choice of $K$ revealed that, under the baseline Transformer decoder, insertion errors rarely occurred due to repetition, particularly in very short or noisy segments. In contrast, the hybrid decoding process mitigates meaningless token repetition by leveraging the complementary strengths of the two decoders. As a result, insertion errors are alleviated to some extent compared to the baseline. These dynamics contributed to differences in WER depending on the value of $K$.

With respect to latency, our results show that the transformer decoder requires significantly more inference time, especially as the output token count increases. This trend is clearly illustrated in Figure~\ref{fig:token_bin_inference_time}, which shows the average inference time for the encoder, transformer decoder, fast decoder, and hybrid decoding, grouped by every 10 output tokens when $K=3$. For the baseline transformer decoder, inference time increases sharply in proportion to the number of output tokens, reflecting its greater computational complexity compared to the fast decoder. By contrast, for the proposed hybrid method, the latency difference is negligible when the output token count is small, but as the output sequence becomes longer, the latency is greatly reduced due to the involvement of the fast decoder.

Consistently across datasets, the encoder module contributes a similar inference time regardless of output length, confirming that decoder computation is the primary contributor to total latency. Critical here is that the proposed hybrid method maintains the same WER as the transformer baseline in-domain, while reducing latency by up to 3.4 times. Even on out-of-domain data, the hybrid method matches the underlying transformer model's WER and far outpaces it in speed. While TDT is even faster, its accuracy drops substantially, especially in the out-of-domain setting.

\begin{figure}[tbp]
\centering
\includegraphics[width=0.95\linewidth]{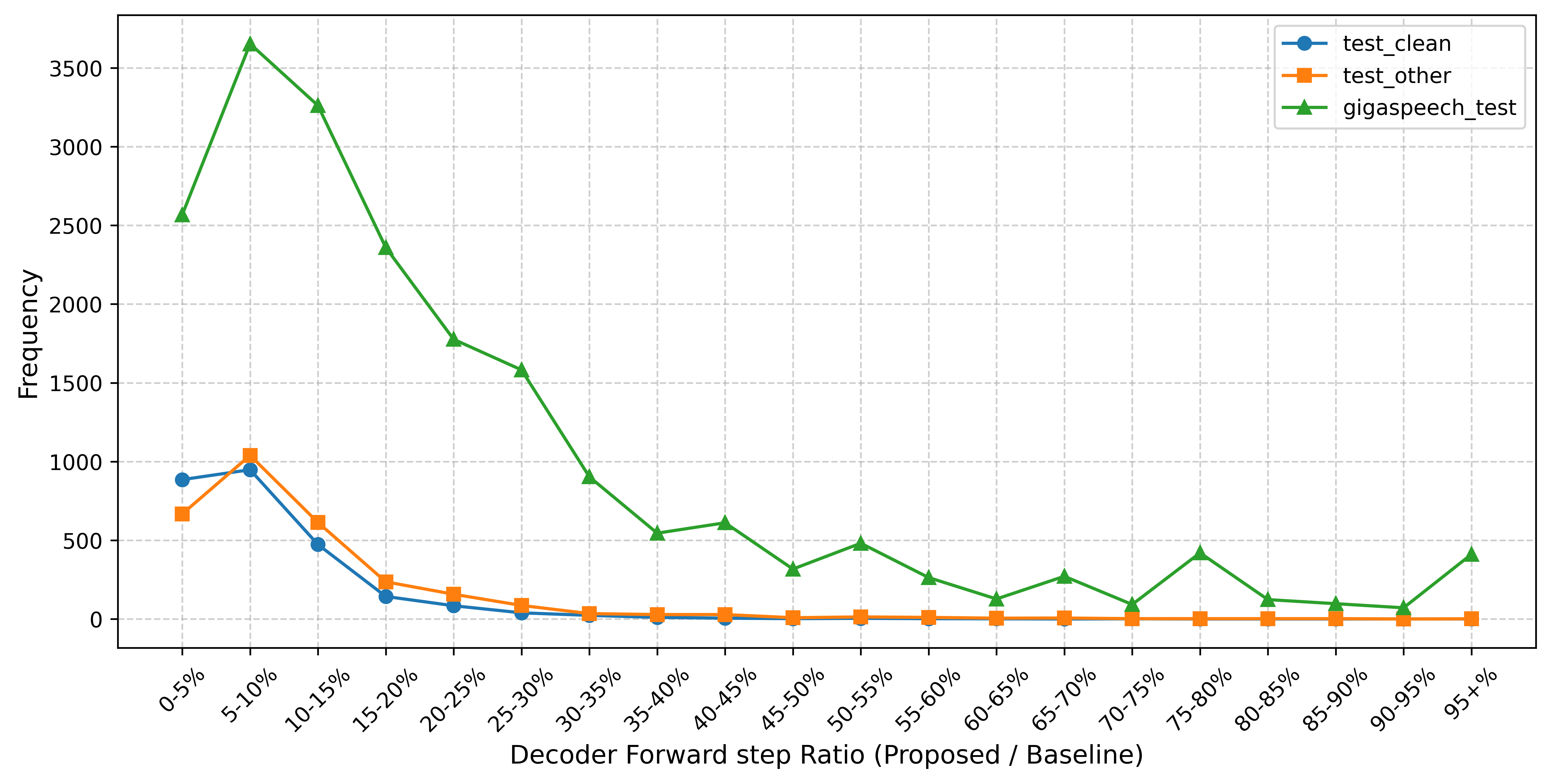}
\caption{Distribution of the ratio of transformer decoder forward steps for the proposed hybrid decoding ($K=3$) compared to the baseline, segmented by output length quantiles. A lower ratio indicates that substantially fewer forward steps are required by the hybrid method, especially for longer sequences, leading to improved inference efficiency. The majority of samples achieve a significant reduction in computational workload for the transformer decoder.}
\label{fig:forward_step_ratio}
\end{figure}

Figure~\ref{fig:forward_step_ratio} illustrates the number of utterances for each ratio of transformer decoder forward steps between the baseline and the proposed method at $K=3$. A lower ratio indicates that the fast decoder significantly reduces the number of transformer decoder forward steps, while a higher ratio implies the opposite. In LibriSpeech, where the fast decoder performs highly effectively, 90\% of the samples are contained within the 30\% ratio bin. In GigaSpeech, the majority of samples fall within the 45\% bin, primarily due to recognition errors from the fast decoder. These results demonstrate both the efficiency and the robustness of our proposed approach.

\begin{figure}[tbp]
\centering
\includegraphics[width=0.95\linewidth]{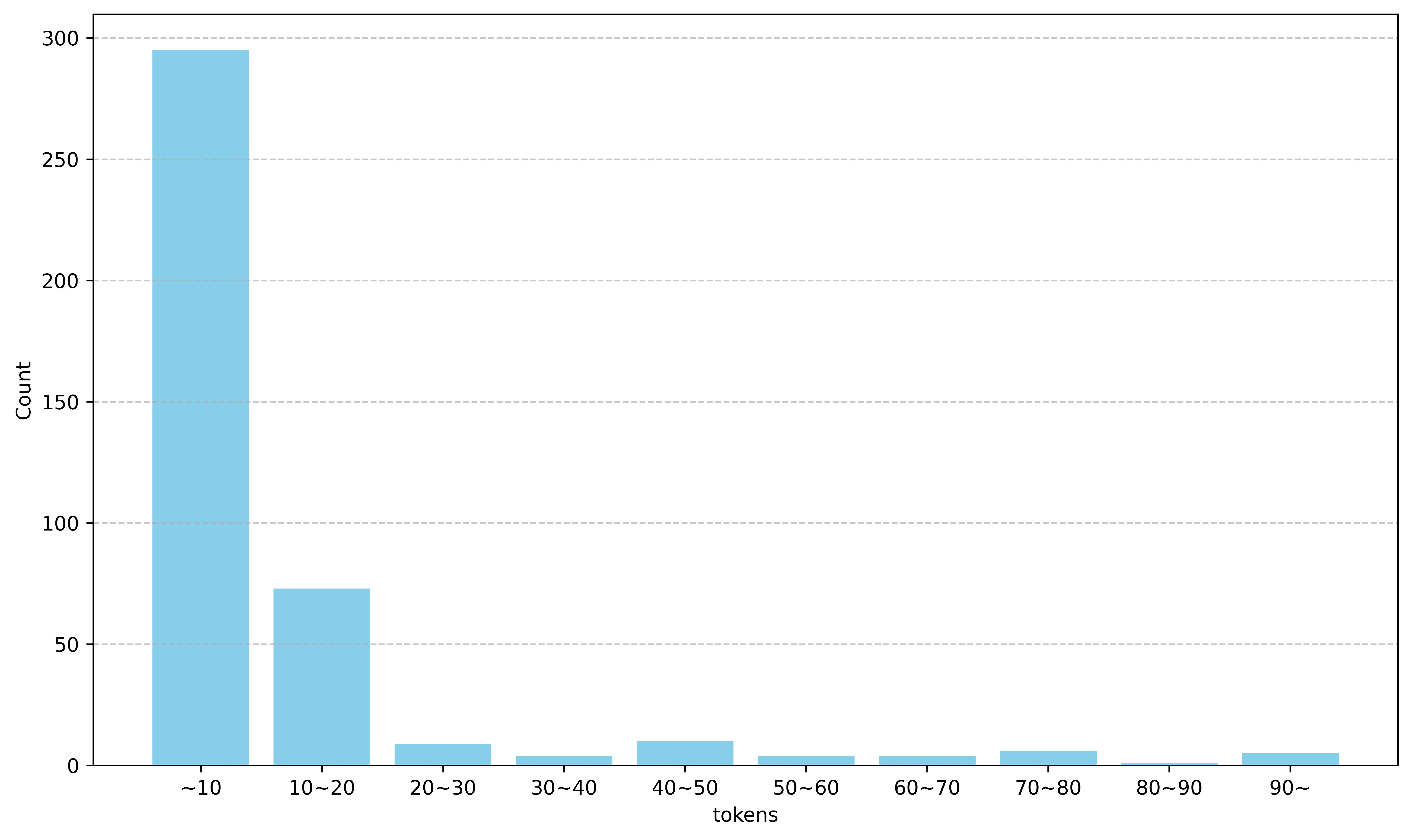}
\caption{Histogram of output sequence lengths for utterances where the ratio of transformer decoder forward steps (hybrid to baseline) is 95\% or higher. The figure shows that the majority of cases with minimal efficiency gain occur for very short utterances (fewer than 10 tokens), confirming that the hybrid approach is most beneficial for longer sequences and that overall latency remains low for short sequences regardless of decoding method.}
\label{fig:forward_step_ratio_95+}
\end{figure}

Figure~\ref{fig:forward_step_ratio_95+} shows the frequency distribution by token count for utterances in which the proposed and baseline methods yield similar numbers of forward steps (i.e., ratio $\geq$ 95\%). Notably, 71.5\% of these cases correspond to utterances with fewer than 10 tokens, suggesting that the proposed method may be less efficient for short utterances where recognition errors from the fast decoder occur. Nevertheless, since the latency for short sequences is generally tolerable, and further gains in efficiency are expected as the fast decoder improves, this limitation is unlikely to be a significant concern.

\section{Conclusion}
In this work, we proposed a novel hybrid decoding approach for attention-based encoder-decoder (AED) speech recognition models, addressing both the inefficiency of fully autoregressive decoders and their susceptibility to certain error modes such as repetition. By introducing a lightweight fast decoder attached to the pretrained encoder and employing a two-pass correction process via teacher-forced verification and localized patching, our method achieves a substantial reduction in inference latency without compromising recognition accuracy. Experiments across both in-domain (LibriSpeech) and out-of-domain (GigaSpeech) datasets show that our approach matches or surpasses the word error rates of strong transformer-based baselines, while delivering more than a twofold speedup in decoding. The efficiency gains of our approach become even more pronounced for longer output sequences, demonstrating its scalability and practicality for real-world ASR systems.

\section{limitations}
Although the proposed method demonstrates substantial improvements in inference efficiency and maintains recognition accuracy, several limitations remain. First, our approach is currently applied only to greedy decoding. Incorporating the framework into beam search decoding could potentially yield further gains in recognition performance and robustness. Second, the current design enforces a unidirectional correction flow, where the process always proceeds from the fast decoder to the Transformer decoder. Enabling bidirectional interaction between the decoders—such as feeding corrections or uncertainty estimates back to the fast decoder—might further enhance error correction and robustness. Third, comparisons with other accelerated decoding methods, such as early exit and speculative decoding, were not conducted due to the unavailability of training data used for the backbone model. Nevertheless, it is worth emphasizing that our proposed method is complementary to existing transformer decoder acceleration techniques and can be seamlessly integrated with them. Exploring these extensions remains an avenue for future work.

\end{document}